\newcommand{\refFig}[1]{Fig.~\ref{#1}}

\newcommand{\change}[1]{{\color{black}{#1}}}

\documentclass[journal,transmag,comsoc,peerreview,final]{IEEEtran}
\IEEEoverridecommandlockouts
\usepackage{cite}
\usepackage{url}

\usepackage[cmex10]{amsmath}
\usepackage{mdwmath}
\usepackage{mdwtab}
\usepackage{amsfonts,amsmath,color,xcolor,amssymb,amsxtra,graphicx,floatflt}
\usepackage{subfigure,tabularx}
\usepackage{booktabs}
\usepackage{multirow}
\usepackage{nth}
\usepackage{epstopdf}

\makeatletter
\let\oldfootnote\footnote
\def\footnote{\@ifstar\footnote@star\footnote@nostar}
\def\footnote@star#1{{\let\thefootnote\relax\footnotetext{#1}}}
\def\footnote@nostar{\oldfootnote}
\makeatother

\begin{document}


\title{Multi-access Edge Computing:\\ The driver behind the wheel of 5G-connected cars}
\author{Fabio Giust,~\IEEEmembership{Member,~IEEE}, Vincenzo Sciancalepore,~\IEEEmembership{Member,~IEEE}, Dario Sabella,\\Miltiades C. Filippou,~\IEEEmembership{Member,~IEEE}, Simone Mangiante, Walter Featherstone, Daniele Munaretto\\
\thanks{\textit{F. Giust and V. Sciancalepore are with NEC Laboratories Europe GmbH, 69115 Heidelberg, Germany.}}
\thanks{\textit{Emails: fabio.giust@neclab.eu, vincenzo.sciancalepore@neclab.eu}}
\thanks{}
\thanks{\textit{D. Sabella and M. Filippou are with Intel Deutschland GmbH, 85579, Neubiberg, Germany.}}
\thanks{\textit{Emails: dario.sabella@intel.com, miltiadis.filippou@intel.com}}
\thanks{}
\thanks{\textit{S. Mangiante is with Vodafone Group R$\&$D, Newbury, UK.}}
\thanks{\textit{Email: simone.mangiante@vodafone.com}}
\thanks{}
\thanks{\textit{W. Featherstone is with Viavi Solutions, Newbury, UK.}}
\thanks{\textit{Email: walter.featherstone@viavisolutions.com}}
\thanks{}
\thanks{\textit{D. Munaretto is with Athonet s.r.l., 36050 Bolzano vicentino VI, Italy.}}
\thanks{\textit{Email: daniele.munaretto@athonet.com}}
}

\ifCLASSOPTIONpeerreview
	\markboth{Submitted to IEEE Communications Standards Magazine}%
	{}
\else
	\markboth{Submitted to IEEE Communications Magazine}%
	{Han \MakeLowercase{\textit{et al.}}: D2D-Based Grouped RA to Mitigate Mobile Access Congestion in 5G Sensor Networks}
\fi

\maketitle
\setlength{\textfloatsep}{5pt}

\begin{abstract}

The automotive and telco industries have taken an investment bet on the connected car market, pushing for the digital transformation of the sector by exploiting recent Information and Communication Technology (ICT) progress. As ICT developments continue, it is expected that the technology advancements will be able to fulfill the sophisticated requirements for vehicular use cases, such as low latency and reliable communications for safety, high computing power to process large amount of sensed data, and increased bandwidth for on-board infotainment.

The aforementioned requirements have received significant focus during the ongoing definition of the 3GPP 5G mobile standards, where there has been a drive to facilitate vertical industries such as automotive, in addition to providing the core aspects of the communication infrastructure. Of the technology enablers for 5G, Multi-access Edge Computing (MEC) can be considered essential. That is, a cloud environment located at the edge of the network, in proximity of the end-users and coupled with the service provider's network infrastructure. Even before 5G is rolled out, current mobile networks can already target support for these challenging use cases using MEC technology. This is because MEC is able to fulfill low latency and high bandwidth requirements, and, in addition, it lends itself to be deployed at the vertical industrial sector premises such as road infrastructure, air/sea ports, smart factories, etc., thus, bringing computing power where it is needed most. 

This work showcases the automotive use cases that are relevant for MEC, providing insights into the technologies specified and investigated by the ETSI MEC Industry Specification Group (ISG), who were the pioneer in creating a standardized computing platform for advanced mobile networks with regards to network edge related use cases.

\end{abstract}

\begin{IEEEkeywords}
V2X, 5GAA, ITS, ETSI MEC, edge computing, orchestration, C-RAN, OpenAPI, API.
\end{IEEEkeywords}

\IEEEpeerreviewmaketitle

\thispagestyle{empty}

\section{Introduction}

\IEEEPARstart{T}{he} automotive sector has been suffering for several years from the global economic crisis. The overall picture showed a less attractive (and remunerative) segment to invest with respect to other industries. With the advent of advanced communication technologies, the automotive area has recovered exhibiting industry profits much higher and expected to grow even faster in the next years. However, the only few winners will be the ones able to put new capital outlays for implementing the digital transformation into their service offering and product portfolio, embracing technologies such as connected cars and autonomous driving~\cite{road2020}.

To assist this large business expansion, the telco industry has included the support to Vehicular-to-Everything (V2X) communication already defined in 3GPP Release 14 LTE specifications. This would find its complete realization with the next-generation of mobile networks (5G) deployment envisioned as a \emph{network design revolution} that accommodates new extreme service requirements while enabling Original Equipment Manufacturers (OEMs) and automotive vertical segments to improve their return on capital.
For this purpose, the telco and the automotive worlds gathered and different organizations emerged, such as the 5G Automotive Association (5GAA)\footnote{http://5gaa.org/} and the European Automotive Telecom Alliance (EATA)\footnote{European Automobile Manufacturers Association (ACEA): "37 leading companies join forces in European Automotive-Telecom Alliance", available at http://www.acea.be/press-releases/article/37-leading-companies-join-forces-in-european-automotive-telecom-alliance} 
aiming at quickly developing new use cases and technologies, and bringing them to research communities as well as to Standards Developing Organizations (SDOs).

This phenomenon introduces new digital features on the vehicles that require high-computational and complex software, very low delay in the data exchange and ubiquitous connectivity.
These requirements are expected to be fulfilled by the recent trend in developing Multi-Access Edge Computing (MEC) solutions, i.e., bringing a powerful cloud computing environment at the edge of the network, e.g., within the Radio Access Network (RAN) premises of a mobile carrier thereby fully exploiting the bandwidth and the available reduced communication latency.
In late 2014, the European Telecommunications Standards Institute (ETSI) launched the MEC Industry Specification Group (MEC ISG) to develop a standardized environment for efficient and seamless integration of applications from ICT suppliers, service providers, and third-parties (e.g., car OEMs) across multi-vendor computing platforms at the edge of mobile networks.
Starting with use-case-driven requirements, the ISG has defined a reference architecture in Group Specifications (GS) MEC 003 and a set of APIs for MEC key-interfaces.
These include specifications relating to the essential functionality of $i$) the application-enabled platform (API framework), $ii$) specific service-related APIs and $iii$) management and orchestration-related APIs\,\footnote{ETSI MEC documents, whether already published or under way, are only mentioned throughout the paper by their reference number without a bibliographic entry. We refer the reader to all published specifications at the following ETSI MEC web page: http://www.etsi.org/techno4logies-clusters/technologies/multi-access-edge-computing}. 
Such APIs are designed to be application-developer-friendly and easy-to-implement so as to stimulate innovation and foster the development of automotive-oriented applications~\cite{giust_etsi_mec}.

Recently, the ETSI MEC ISG has expanded the scope of its activities to include additional access technologies besides cellular and support Internet-of-Things (IoT) deployments with low-energy support~\cite{energy_iotj2016} and connected cars, thus, updating the MEC term into \emph{Multi-Access} Edge Computing\footnote{http://www.etsi.org/news-events/news/1180-2017-03-news-etsi-multi-access-edge-computing-starts-second-phase-and-renews-leadership-team}. This aimed at strengthening the engagement with OEMs and service providers as the stakeholders that exploit MEC for their added-value product propositions.

\change{In light of such considerations about V2X and MEC, in this paper we thoroughly study the effort of ETSI ISG to address the automotive use cases and we shed the light on advantages and technical challenges while addressing the main advanced vertical requirements.
In particular, we focus on the use of MEC technology in V2X scenarios, the definition and implementation of MEC APIs, the application of the MEC concept in current network deployments and next-generation networks, the smart orchestration of MEC resources and its relationship with the novel network slicing paradigm as the key-enabler of 5G automotive applications.
Finally, we illustrate practical issues and technical aspects inherent to the deployment of the ETSI MEC technology while showing how ETSI MEC ISG will tackle those in the next future.}

\section{MEC-empowered V2X use-cases}
\label{sec:v2x}
\begin{table}[t!]
\caption{V2X use cases and relevance for MEC}
\label{tab:use_cases}
\scriptsize
\centering
\begin{tabular}{|l||l|c|}
\hline
\textbf{Use cases} & \textbf{Description} & \textbf{MEC Relevance}\\
\hline
\textit{Intersection Movement} & Warn Host Vehicle (HV) driver & $\uparrow\uparrow\uparrow$ High\\
\textit{Assist}	& of collision risk through an & \\
				 & intersection. & \\
\hline
\textit{Software Updates} & Deliver and manage automotive & $\uparrow$ Mid\\
				 & software updates. & \\
\hline
\textit{Real-Time Situational} & Alert HV driver of hazardous & $\uparrow\uparrow\uparrow$ High\\
\textit{Awareness \& High} &  (e.g., ice) road conditions ahead. & \\
\textit{Definition Maps} &  & \\
\hline
\textit{See-Through} & Provide HV driver with a video & $\uparrow\uparrow\uparrow$ High\\
			& stream of the view in front of & \\
            & the Remote Vehicle (RV) intended & \\
            & to be passed using the oncoming & \\
            & traffic lane. & \\
\hline
\textit{Cooperative Lane} & Signaling by the HV to, at least, & $\uparrow\uparrow$ High\\
\textit{Change (CLC)}		  & one RV of the HV driver's intention& \\
\textit{of Automated Vehicles} & to change lane to the RV's lane.& \\
\hline
\textit{Vulnerable Road} & Detects VRUs in the vicinity & $\uparrow\uparrow\uparrow$ High\\
\textit{User (VRU) Discovery} & of HVs and warns their drivers. & \\
\hline
\end{tabular}

\end{table}

\begin{figure*}[!t]
\centering
\includegraphics[width=2\columnwidth]{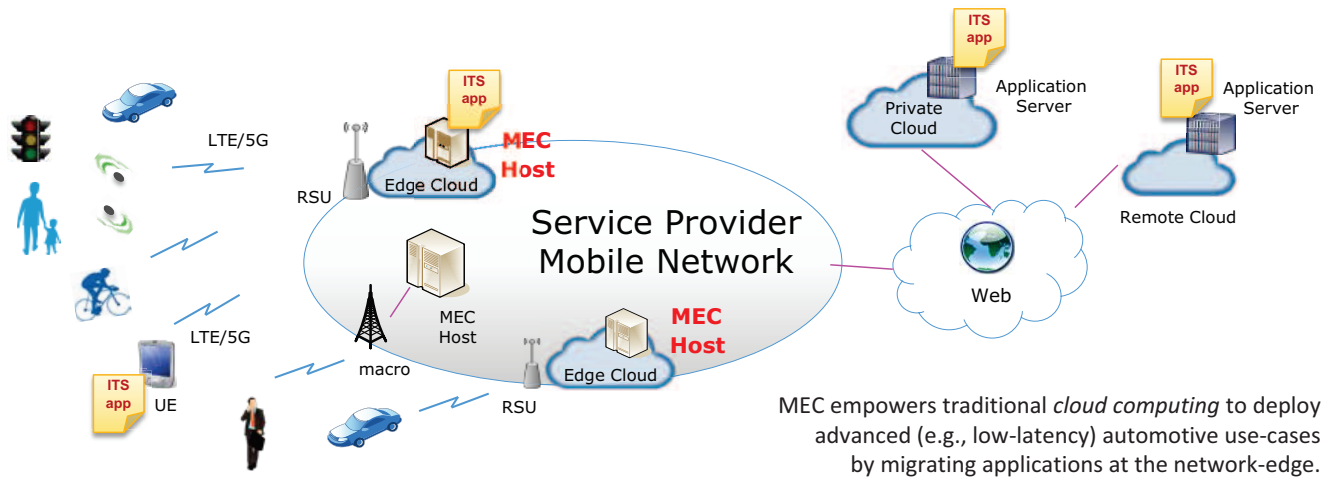}
\caption{Example of an automotive scenario wherein 5G empowers an MEC deployment.}
\label{fig:v2x}
\end{figure*}

The automotive sector has unveiled its importance as a key-driver towards the adoption of MEC-based solutions among the vertical segments involved in the 5G ecosystem~\cite{5gaa_mec}. This is driven by the fact that the main communication paradigm, namely Vehicle-to-Infrastructure (V2I) communication, is envisioned to require a \emph{very low-latency} environment for a significant proportion of the related use-cases \change{($<10\,$ms according to~\cite{mec_cran})}. Such a capability is offered by MEC, which facilitates the direct exchange of V2X related information between (mobile) nodes via the underlying communication network.
Further evidence of the importance of this synergy is supported by many experimental activities\,\footnote{Continental, Deutsche Telekom, Fraunhofer ESK, and Nokia Networks: ``Digital A9 Motorway Test Bed'', available at http://www.prnewswire.com/news-releases/continental-deutsche-telekom-fraunhofer-esk-and-nokia-networks-showcase-first-safety-applications-at-digital-a9-motorway-test-bed-543728312.html} and trials on automotive use cases\,\footnote{Press release: ``Comba Telecom and ASTRI demonstrate Mobile Edge Computing System (MEC) with Vehicle-to-Everything (V2X) system prototype at Mobile World Congress 2017 in Barcelona'' available at http://markets.businessinsider.com/news/stocks/Comba-Telecom-and-ASTRI-demonstrate-Mobile-Edge-Computing-System-MEC-with-Vehicle-to-Everything-V2X-system-prototype-at-Mobile-World-Congress-2017-in-Barcelona-1001792754} as result of the collaboration between stakeholders, ranging from car makers and OEM suppliers to purely telco operators.
Table~\ref{tab:use_cases} summarizes a number of key V2X use cases focused towards safety and considered essential for connected and Autonomous Driving (AD) vehicles, showing the level of relevance for MEC systems according to industry consortia, such as the 5GAA.

Infotainment is also expected to offer considerable business opportunities, where, according to recent estimates,\,\footnote{Intel webinar: ``Intel 5G Connected Vehicles Webinar'', available at \url{https://pages.questexweb.com/Intel-Registration-012918.html?source=intel }} a new $\$$7 trillion worth passenger economy will emerge when drivers become passengers, freeing up to $300$ hours per year they typically spend behind the wheel. 5G, in combination with MEC technologies, is expected to offer the capability to deliver entertainment media on-board, as the ecosystem transitions from drivers to riders. 

In summary, both infotainment and AD use-cases can be significantly impacted by MEC system deployments. Focusing on these use cases, many devices (including cars, motorcycles, bicycles, pedestrians, etc.) need to be fully connected, communicate and exchange information related to traffic conditions, surrounding environment, maps, but also generate, share and retrieve diverse content, such as video streaming, Augmented Reality/ Virtual Reality (AR/VR) services or simple messaging operations. \change{As a consequence, the role of \emph{edge computing} appears to be essential to provide fast access to and distribution of information of local interest, where the location of the users is fundamental, such as safety alarms, environmental characteristics like the proximity of other vehicles (e.g., \textit{See through} use case) and obstacles (e.g., \textit{Real-time situational awareness} use case). MEC is crucial also to guarantee a preferred environment for local pre-processing of data, for instance to reduce the consumed backhaul bandwidth, as well as the possibility to host---in close proximity to user devices---Intelligent Transport System (ITS) applications, such as the 3GPP-defined V2X application server~\cite{23.285}.}

Fig.~\ref{fig:v2x} shows an illustrative example of a reference V2X system overlaid by an MEC deployment, where cars communicate with Road Side Units (RSUs) and base stations. MEC hosts are co-located with those RAN elements so as to allow the consumption of ITS applications by exploiting the distributed computing environment, including terminals, the edge cloud, as well as private and remote clouds.

The MEC system facilitates Information Technology (IT) cloud capability at the edge of the network, and its access-agnostic characteristics guarantee a smoother deployment independently of the underlying communication network (e.g., 3GPP LTE, its evolution, 5G, or other non-3GPP access technologies). In addition, the MEC system is able to expose a set of standardized interfaces to such applications, which permit developers and car OEM suppliers to implement ITS services in an inter-operable (and flexible) way across heterogeneous access networks, mobile operators and vendors.
Indeed, this represents the \emph{added-value} brought by ETSI MEC (as a unique standard for edge computing): Standardized interfaces, namely MEC APIs, allow the interoperability across different stakeholders in the automotive system, as described in the next section.

The engagement of ETSI MEC ISG with vehicular networking has been phased into two stages.
The first part consists in the aforementioned collection of V2X use cases and the definition of new technical requirements of the MEC system.
Such activity is documented in the Group Report (GR) MEC 022, \textit{Study on MEC support for V2X use cases}.
Therein, it emerged that some of the most important requirements, commonly identified by automotive stakeholders, stem from multi-Mobile Network Operator (MNO) and multi-OEM scenarios: in principle, an ITS operator may be interested in offering an ITS service to its own users across different communication networks, and, potentially, involving a set of vehicles from different manufacturers.

As a follow-up stage, future work is envisaged to include new normative activities in order to close the gap identified by the study, e.g., by defining
standardized interfaces to overcome potential interoperability issues in the earlier depicted multi-MNO multi-OEM scenario.

\begin{figure}[t]
\centering
\includegraphics[width=\linewidth]{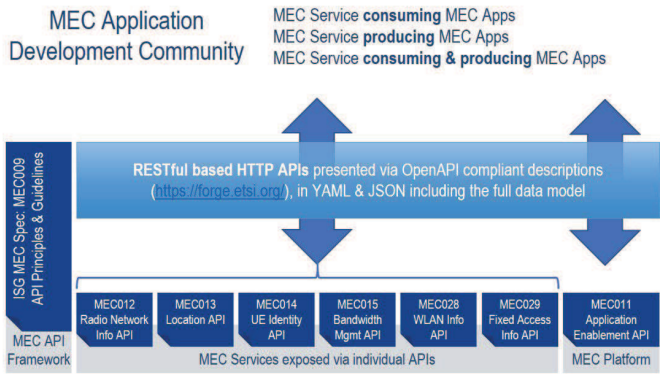}
\caption{MEC framework implementing REST-based APIs.}
\label{fig:mec-apis}
\end{figure}

\section{MEC APIs for added value services}
\label{sec:mec_apis}

The MEC APIs mentioned in the previous section are designed to serve
a wide plethora of 5G use cases, among which those belonging to automotive scenarios. Their definition represents a key activity within the MEC ISG; the diagram in \refFig{fig:mec-apis} illustrates the associated API framework, including the published and under-development APIs.


Among the services, the \emph{Radio Network Information Service (RNIS)}, specified in GS MEC 012, is a key feature offered by MEC, as it provides actual RAN information relating to the User Equipments (UEs) in the network. 
Future API refinements aim at providing V2X applications with predictions on the Quality-of-Service (QoS) performance of the V2I link; this is a fundamental feature for car OEMs, not only to improve reliability, but also to support fast (and dynamic) decisions in the next road segment, where, for instance, vehicles may switch to a safe off-line mode if the QoS forecast is poor.
In combination with RNIS, \emph{MEC Location Service (LS)}, specified in GS MEC 013, is a powerful tool enabling applications to exploit user proximity information, e.g., to retrieve and monitor the list of users connected to a particular cell or access point.
This service is essential for V2X applications to estimate the location of vehicles as well as of pedestrians, e.g., to improve their safety.
Additionally, the \emph{UE Identity Service}, specified in GS MEC 014, allows applications to trigger user-specific traffic rules on the MEC platform by e.g., steering traffic to a local network.
The final specified service API is the \emph{Bandwidth Management Service}, described in GS MEC 015, which offers applications the ability to reserve networking resources at the host, thereby ensuring Quality-of-Experience (QoE) requirements can be achieved.

By expanding its scope to encompass \emph{Multi-access}, the ISG has initiated the development of additional service APIs.
First, the \textit{Wireless Local Area Network (WLAN) Information API}, specified in GS MEC 028, which emulates the 3GPP mobile access focused RNIS, but for WLAN deployments. Second, the \textit{ Fixed Access Information API}, GS MEC 029, which has a wider remit to cover Fibre, Cable, xDSL, and Point-to-Point Fibre Ethernet access to MEC. The goal is to develop a generic API proving access network related information for the multitude of fixed access technologies.
ETSI MEC pursues also additional service APIs, one targeting specifically the V2X use cases with the aim of facilitating V2X interoperability in a multi-vendor, multi-network and multi-access environment\,\footnote{Note that no GS document drafts are available at the time of writing.}.

Since MEC services (current and future) constitute a key added value, a service handling functional block in the platform is of paramount importance in order to allow applications discovering or exposing them.
The purpose of the \textit{Mobile Edge Platform Application Enablement API}, described in GS MEC 011, is precisely to specify the set of environmental interfaces exposed by the MEC platform enabling applications to discover the services they wish to consume, or to register and advertise the services they intend to offer.


A key objective for the ISG from its inception has been third party developer engagement and adoption of open-source principles, which illustrates a shift away from the ETSI more traditional telco-centric approach to standards development.
To this end, all of the published APIs have been made available via ETSI forge site\footnote{https://forge.etsi.org/}, where they are presented via OpenAPI Specification (OAS)\footnote{https://github.com/OAI/OpenAPI-Specification} compliant descriptions. These descriptions enable interactive documentation of the APIs, allowing users to interact with each of the services to better understand their operation and the capabilities provided.
Third party tooling is also readily available for auto generation of client and server stubs (multiple languages supported, e.g., Node.js, Java, Go), API testing and automated OAS compliance checking. In this manner, the accessibility of the service APIs is enhanced, with the aim of creating a larger and wider edge focused ecosystem.
To further that goal, a framework to support and encourage MEC Hackathons has also been established\footnote{https://mecwiki.etsi.org}, with the primary aim of inspiring innovation, but also to promote the MEC standards and seek associated feedback.
This could represent an intermediate step towards trialling MEC-based V2X use cases, for instance, by implementing the network design proposed in the next section.

\section{Example of a field setup of MEC for V2X: a step forward to 5G}
\label{sec:deployment}
\begin{figure*}[!t]
\centering
\includegraphics[width=2\columnwidth]{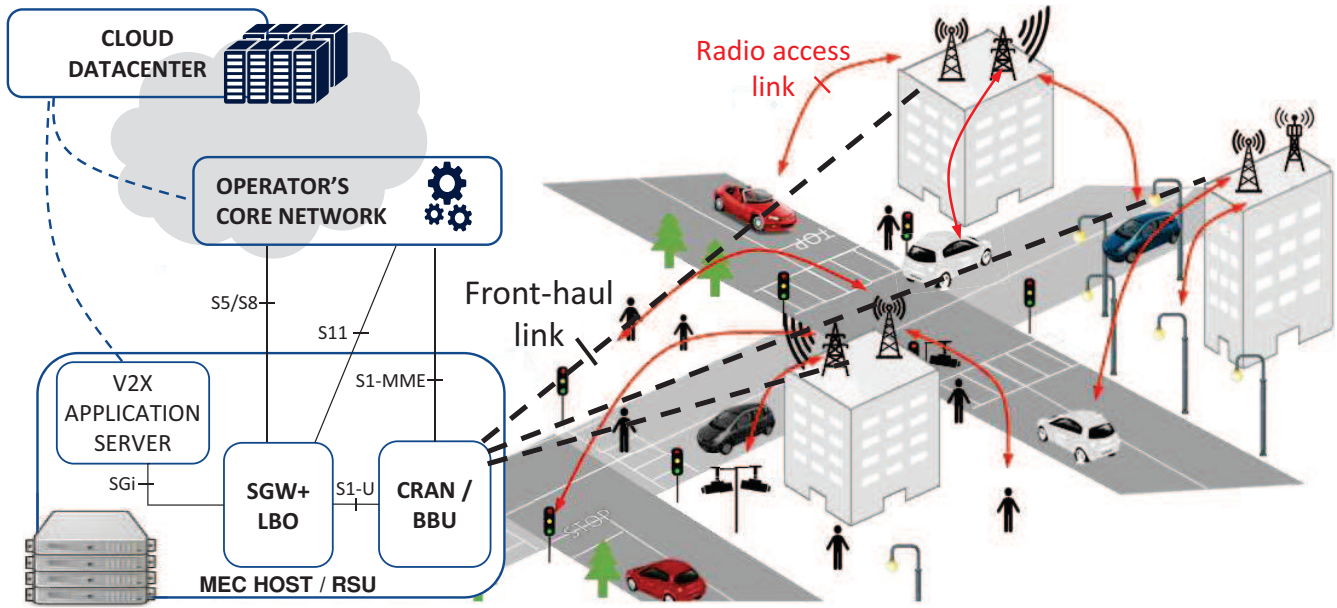}
\caption{Example of a MEC deployment for V2X featuring C-RAN and a 4G core network.}
\label{fig:mec-v2x}
\end{figure*}

The introduced 5G service types, such as e/xMBB (enhanced/extreme Mobile Broad Band), URLLC (Ultra Reliability and Low Latency Communications), and mMTC (massive Machine Type Communications) exhibit performance metrics highly dissimilar or even conflicting each other.
For this reason, in the envisioned coexistence of multiple virtualized networks, namely \emph{network slices}, for instance, tailored to automotive use cases, business plans as well as access and core network architectural upgrades need to progress alongside~\cite{Campolo2017}.
\change{In addition, the transition to a fully capable 5G network is expected to happen by gradually taking over the 4G equipment.
MEC promises to accelerate such transition, being able to fulfill many of the requirements to realize automotive use cases.
In the following, we will provide an example, illustrated in \refFig{fig:mec-v2x}, of how an MEC host elevates the concept of an RSU to another dimension, able to accommodate the latest equipment for vehicular and mobile networking, with no need to deploy strict 5G components.}

Conventionally, the RAN constitutes the main source of MNOs' Capital Expenditure (CAPEX) and Operational Expenditure (OPEX), due to the extensive use of physical network components such as Baseband Units (BBUs), switches and other entities distributed across the sites spanning a given area of deployment.
To tackle the cost-related drawbacks of a highly distributed RAN implementation, the wireless communication ecosystem has recently focused on RAN centralization techniques following a ``phased'' deployment approach. The first step towards realizing this approach was the (partial) centralization of the RAN by means of clustering BBUs, for a number of sites, to a Central Office (CO), so as to enhance performance via Coordinated Multipoint (CoMP) transmissions, however, depending on the quality of used fronthaul interfaces.

Moving a step forward, according to the Cloud-RAN (C-RAN) approach, the various clustered BBUs can form a ``BBU pool'', thus, allowing for a dynamic allocation of processing resources to accommodate the---dynamically changing---performance needs of the different sites (and, even different Radio Access Technologies - RATs). 
Such dynamic resource allocation leads towards more efficient resource coordination, however, still under the constraint of exploiting dedicated hardware entities. To enhance hardware reconfigurability and network agility, especially in view of high mobility scenarios, recent advents in Network Functions Virtualisation (NFV) and Software Defined Networking (SDN) have boosted the potential of a C-RAN deployment by means of virtualizing network functions, thus, enabling General Purpose Processors (GPPs) to be exploited for optimized network operations.

Nevertheless, the end-to-end latency requirements of all the automotive use cases (e.g., the need to periodically exchange Cooperative Awareness Messages - CAM - in a timely manner) are quite stringent to achieve by means of a highly virtualized, albeit ``remote'' network cloud.
On top of that, the need to process data in a proximity-based fashion, towards e.g. enhancing privacy/security, or, depending on the context (e.g., vehicle location, driving conditions) can be hardly satisfied even by means of fractional RAN centralization.
Towards this end, the proximity-oriented deployment of MEC hosts at RSUs, or, collocated with Base Stations (BSs)~\cite{Emara2017}, can be proven useful for efficiently allocating the resources needed to run C-RAN equipment as well~\cite{mec_cran}.

Architecturally renewed by the 5G service types and related key features, the core network shall support new highly demanding services with increased data rate, reduced end-to-end latency, massive connectivity, guaranteed QoS/QoE, higher availability and efficiency.
In this picture, MEC is expected to facilitate the transition to 5G deployments, as it accommodates the co-location of core network functions at edge cloud facilities together with the MEC applications, hence having 5G use cases fulfilled using the 4G technology~\cite{klas2017edge}.
In this context, and according to recent MEC deployment proposals~\cite{mec_4g5g}, the Serving Gateway function can be distributed at the edge and enhanced with an IP-based interface, similarly to the SGi interface of the Packet Data Network (PDN) Gateway
, which selectively breaks out the traffic that needs to remain local, i.e., associated to MEC applications as well as to the aforementioned V2X application server.
Based on this description, the solution is named Serving Gateway-Local Breakout (SGW-LBO).
Similarly to the uplink classifier, it allows the operator to steer the traffic based on different parameters such as Access Point Name (APN), user identities, packet header's 5-tuple, Differentiated Services Code Point (DSCP), while supporting user mobility, charging, lawful interception, and session management operations.
This MEC solution is fully 3GPP standards-compliant and it does not require changes in the standard 3GPP interfaces that would greatly limit the impact on the operations of an MNO. 

Accounting for the above-considerations, it appears straightforward to leverage edge computing facilities within a V2X communications system, while incorporating the following main sets of functionalities: $i$) BBUs of C-RAN for access link termination, $ii$) core gateway functions for session management, routing of local traffic, etc., and $iii$) an inter-operable and programmable computing environment for cloud-based vehicular applications.

\section{Orchestrating edge clouds}
\label{sec:orchestrating}

\begin{figure}[t]
\centering
\includegraphics[width=\columnwidth]{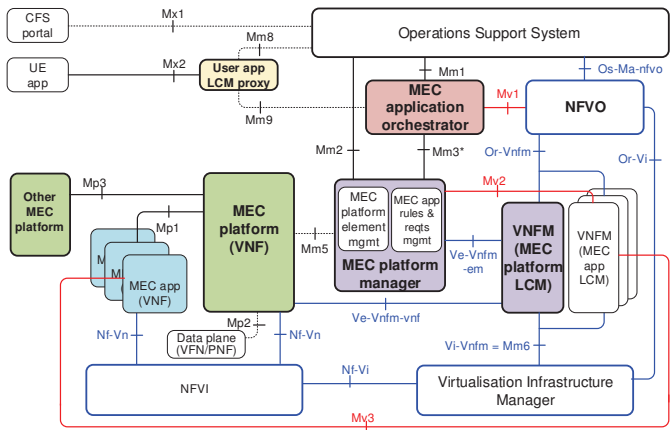}
\caption{MEC architecture in an NFV environment.}
\label{fig:mec-nfv-arch}
\end{figure}

The previous section introduced a deployment scenario where components of MEC, 4G and vehicular networks coexist in the same computing environment at the edge of the network, thereby implicitly assuming some sort of virtualized infrastructure underneath to run such components as software instances.

Such an assumption is motivated by observing that MNOs are currently in the process of virtualising their networks adopting the NFV standard~\cite{nfv002} in order to achieve several benefits including higher scalability, higher service deployment speed, cost reduction, and easy management of multi-vendor assets. ETSI NFV already defines a comprehensive Management and Orchestration framework (MANO)~\cite{nfv-mano} for orchestrating Virtualised Network Functions (VNFs) and integrating them with external resources (e.g., billing systems and customer portals) through standardised interfaces (APIs).

With the introduction of MEC to handle V2X use cases, MNOs will face the challenge of orchestrating all their edge nodes, geographically distributed over several infrastructure locations.
MNOs will look for efficient, open and standard-driven solutions leveraging automation, reusing assets and resources already in place as much as possible (as in the example of the previous section, a site hosting distributed NFV related to C-RAN will be likely used to host edge applications, as well).
We believe that edge orchestration will be mandatory to enable real-time mobility functionalities required by the automotive vertical and V2X use cases.

A three-layer orchestration solution needs to be fully addressed in the MEC deployment:

{\bf Infrastructure:} the management of infrastructure nodes that might differ in size, processing, storage, and latency capabilities; the management of a virtualisation layer (NFVI in ETSI NFV terminology~\cite{nfv002}, e.g., OpenStack\footnote{https://www.openstack.org/}) geographically distributed across multiple clusters.

{\bf MEC platform:} the framework on top of each edge node that enables MEC applications. It must be secure, automatically managed, and integrated with any existing NFV environment. It may provide additional services MEC applications can consume.

{\bf MEC applications:} they require a nested orchestration because they need to consume services, be chained one to another, support user mobility, and be integrated with external clouds. A flexible resource allocation framework within a limited resource pool (by definition) at an edge location must be implemented, taking into account the problem of placement of MEC applications over a number of suitable edge nodes, based on latency and QoS requirements. Although extensively tackled in NFV
, this represents one of the main challenges in orchestrating such a massively distributed architecture, especially when combined to the stringent requirements posed by V2X use cases.

A MEC orchestrator has been properly defined by ETSI MEC 
specifications to fully address the last two mentioned layers, but its role and responsibility need to be further investigated when deployed within a NFV architecture containing its own orchestrator. Specifically, \textit{deployment of MEC in an NFV environment}, described in GR MEC 017, analyses such scenario with the aim of re-using as much as possible the capabilities from the NFV MANO system.
This way, network administrators can jointly manage network functions and applications, thus optimizing the network service creation and orchestration.
Additionally, this allows for monetizing the investment made for the NFV transformation by running MEC applications on top, thus targeting the revenue streams coming from MEC-enabled use cases.

To efficiently cast MEC into the NFV paradigm, the MEC applications and the MEC platform are envisaged as VNFs to be managed and orchestrated by MEC and NFV entities in coordination, provided the correct mapping of MEC functions into NFV functions~\cite{SGSY_CSCN16}.
The legacy MEC management system is kept to handle MEC application specific procedures, such as MEC host selection, traffic and Domain Name System (DNS) rules configuration, and MEC platform management.
Conversely, resource orchestration, MEC application and MEC platform lifecycle management has been devolved to NFV entities.
The ``MEC-in-NFV'' architecture that resulted from the report is depicted in~\refFig{fig:mec-nfv-arch}. The MEC entities (colored with different colors) are shown alongside with the NFV entities (in blue).
In the diagram, black and blue lines draw respectively MEC- and NFV-specific reference points, conveying the MEC or NFV specific interfaces.


From an application standpoint, MEC is the meeting point of telco and IT domains, because third party MEC applications (nowadays developed following a cloud native approach) run inside a highly regulated telecom network. IT and telco worlds differ in development speed, deployment models, security constraints, size of developer communities, and built-in services. As an example, IT and cloud players are betting heavily on micro-services and containers 
for scalable and highly reactive applications, while telco vendors have been working with Virtual Machines (VMs), considered more stable and secure. ETSI MEC is actively engaged with the topic, reflected by the \textit{Study on MEC support for containers}, studied in GR MEC 027.

As MEC nodes perform local breakout of traffic towards external private and public clouds, saving back-haul capacity in the MNOs’ core network, a seamless and elastic integration between MEC applications and the cloud becomes crucial. Such capability may play a key role in attracting to edge computing a large community of developers already used to programming client-server applications in the cloud. V2X application developers will need a reliable connection to the backend systems of car manufacturers and suppliers when dealing with sensitive data or customised processes requiring knowledge from the cloud.
As advised in~\cite{mec_dev_software}, MEC application development demands a shift from traditional cloud models to a 3-tier distributed logic: a smooth transition can be enabled by implementing the correct orchestration, connectivity/reachability, and DevOps tools at the edge.

To summarise, the following emerging challenges in the edge orchestration are exacerbated by V2X applications:
\begin{itemize}
\item On-boarding and testing of MEC applications, especially when provided by developers not aware of NFV and telco procedures;
\item User mobility and consequent MEC application migration between edges within the same MNOs and across different MNOs (e.g., automotive use cases where cars trespass national borders), providing the expected service continuity;
\item Common API exposure across multiple network operators for easier MEC application development and broader adoption;
\item User-generated actions and requests (e.g., a user may decide to deploy and run his application instance in a specific edge node);
\item Security of all the distributed orchestration mechanisms and the hosted MEC applications. A powerful and intelligent monitoring system should be orchestrated at the edge nodes in order to detect threats and react locally.
\end{itemize}

\section{An ETSI-compliant MEC slice tailored to V2X}
\label{sec:slicing}

\begin{figure}[t]
\centering
\includegraphics[width=\columnwidth]{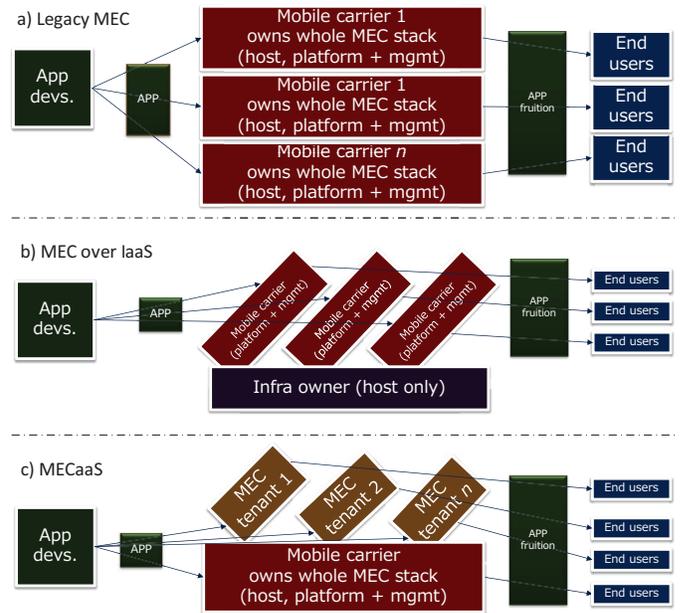}
\caption{Different sharing models of MEC computational resources.}
\label{fig:mec-slicing}
\end{figure}

Considering the MEC-based V2X use case as explained in GR MEC 022, multi-operator support exhibits the most challenging aspect that must be addressed in the upcoming MEC specifications. In particular, a potential solution has emerged recommending the usage of a shared MEC platform among different operators. This calls for a need to support multi-tenancy of MEC facilities e.g., enabled by the network slicing paradigm, requiring a certain effort from the standardisation body. ETSI MEC has already embarked on such effort, and preliminary results are the object of GR MEC 024, namely \textit{MEC support for network slicing}.

Different sharing models can be designed to proper address network slicing in this environment: $i$) static MEC resource split (\emph{Legacy MEC}), $ii$) shared MEC capability over Infrastructure as a Service (IaaS) and $iii$) MEC as a Service (\emph{MECaaS}), as depicted in~\refFig{fig:mec-slicing}.

Legacy MEC represents the legacy model envisioned by today's MEC architecture. Therein, service providers (depicted as boxes in the middle of ~\refFig{fig:mec-slicing}a) such as mobile carriers own the whole MEC system, comprising the infrastructure (i.e., the MEC host), the computing platform and its management system.
Application providers deliver their service packages to the service provider that enables its fruition by end users.

Nevertheless, under some circumstances such as in common V2X scenarios, the service provider is not the owner of the infrastructure, but it may rent the necessary resources to run the MEC platform and management entities following the IaaS model (see ~\refFig{fig:mec-slicing}b).
In this case, the IaaS provider facilitates management tools to enable the MEC orchestrator (owned by the service provider, now a tenant of the MEC infrastructure) to interact with the resources underneath, e.g., via a virtualization infrastructure manager or its abstraction.

Finally, the more interesting and suitable model addressing V2X use cases is proposed in ~\refFig{fig:mec-slicing}c, where a service provider owns the whole MEC stack and it offers part of the totality of the MEC system to tenants (e.g., the automotive industry).
Such exposure and allocation procedures are implemented through a brokering process that directly interacts with network tenants, checks resource availability by means of admission control policies, and grants access to tenant management systems~\cite{ZGS_WCNC18}.
In this way, automotive verticals might seamlessly orchestrate MEC resources, deploy own applications without sharing the source code (or sensible information), and configure policies and rules that best fit the service requirements, in accordance with the rental agreement. 

\section{Concluding remarks}
\label{sec:conclusions}

\change{In this paper, we have explored the Multi-access Edge Computing (MEC) technology, and demonstrated its suitability to underpin V2X use cases, being in facts one of the most promising technology enablers.
%
In particular, we have focused on the ability of MEC to create a standard computing environment able to open to third party's applications the edge of the network, and in turn, to support advanced requirements such as low communication latency thereby posing the technology under the careful consideration of emerging industrial associations that gather stakeholders from the telco and automotive worlds, such as the 5GAA.

We have detailed $i$) the initial MEC framework and reference architecture with the first release of service APIs and management interface specifications and $ii$) the ETSI MEC's engagement with the automotive sector, as per the effort to investigate technical enhancement to support V2X use cases. Such improvements range over a variety of aspects, from the definition to dedicated APIs, to the incorporation of the results from other standardization groups, such as ETSI NFV. Finally, we have shed the light on the new technology trends, such as C-RAN and network slicing, in order to make MEC a future-proof investment able to pave the way towards 5G automotive use cases.}

\bibliographystyle{IEEEtran}
{
\bibliography{references}
}
\vspace{0.5cm}

\end{document}